\documentclass[x11names]{llncs}
\usepackage[utf8]{inputenc}
\usepackage[T1]{fontenc}
\usepackage{csquotes}
\usepackage[english]{babel}
\usepackage{syntax}
\usepackage[table,xcdraw]{xcolor}
\usepackage[unicode=true]{hyperref}
\usepackage{xspace}
\usepackage{amsmath, amssymb}
\usepackage{pdfcomment}

\usepackage{graphicx}
\usepackage{caption, subcaption}
\usepackage{pdfpages}
\usepackage{wrapfig}

\usepackage[backend=biber, maxbibnames=99, firstinits=true]{biblatex}
\usepackage{listings}
\usepackage{lstautogobble}
\lstdefinelanguage{ACSL}{%
	literate=
		{==}{{$==$}}2
		{==>}{{$\Rightarrow$}}1
                {P_M_1}{{$P_{M_1}$}}3
		{integer\ i}{{i$\,\in \mathbb{Z}\,$}}4
		{integer\ j}{{j$\,\in \mathbb{Z}\,$}}4
		{integer\ k}{{k$\,\in \mathbb{Z}\,$}}4
		{unsigned\ i}{{i$\,\in \mathbb{N}\,$}}4
		{unsigned\ j}{{j$\,\in \mathbb{N}\,$}}4
		{unsigned\ k}{{k$\,\in \mathbb{N}\,$}}4
		{unsigned\ page}{{page$\,\in \mathbb{N}\,$}}7
		{\\forall}{{$\forall$}}1
		{\\exists}{{$\exists$}}1
		{integer}{{$\mathbb{Z}$}}1
		{real}{{$\mathbb{R}$}}1
		{&&}{{$\wedge$}}1
		{||}{{$\vee$}}1
		{!=}{{$\neq$}}1
		{<}{{$<$}}1
		{<=}{{$\le~$}}1
		{>}{{$>$}}1
		{>=}{{$\ge~$}}1
		{<==>}{{$\Leftrightarrow$}}1
	,
	morekeywords={
		assert,assigns,assumes,axiom,axiomatic,behavior,behaviors,
		boolean,breaks,complete,continues,decreases,disjoint,ensures,
		exit_behavior,ghost,global,inductive,invariant,lemma,logic,loop, meta,
		model,predicate,relational,reads,requires,sizeof,strong,struct,terminates,
		type,union,variant,uchar,byte,typically,\\result,\\old,\\at,\\valid,
		\\separated,\\nothing,Pre,\\sum,\\numof,\\from,meta,\\writing,
		\\reading,\\calling,\\strong\_invariant,\\weak\_invariant,
		\\written, \\read, \\called
	},
	alsoletter={\\},
	moredelim={*[s]{/*@}{*/}},
	moredelim={*[l]{//@}},
}
\lstnewenvironment{ACSL}
	{\lstset{language=ACSL}}
	{\smallskip}
\lstnewenvironment{CACSL}
	{\lstset{language=C,alsolanguage=ACSL}}
	{\smallskip}

\lstdefinestyle{CACSLstyle}{language={[ANSI]C},%
	alsolanguage=ACSL,
}

\newcommand{\acsl}[1]{\texttt{\textbackslash#1}}

\newcommand{\meta}{\textsc{MetAcsl}\xspace}
\newcommand{\framac}{\textsc{Frama-C}\xspace}
\newcommand{\Wp}{\textsc{Wp}\xspace}

\newif\iflong
\newif\ifshort
\longtrue %
\iflong\shortfalse\else\shorttrue\fi
\newcommand{\shortv}[1]{\ifshort {#1} \fi}
\newcommand{\longv}[1]{\iflong {#1} \fi}

\let\llncssubparagraph\subparagraph
\let\subparagraph\paragraph
\usepackage[compact]{titlesec}
\let\subparagraph\llncssubparagraph
\titlespacing*\section{0pt}{1ex plus 0.3ex minus 1ex}{1ex plus 0.3ex minus 1ex}
\titlespacing*\subsection
{0pt}{0.5ex plus 0.5ex minus 0.2ex}{0.5ex plus 0.2ex minus 0.5ex}
\usepackage[firstpage]{draftwatermark}
\SetWatermarkText{\hspace*{5.7in}\raisebox{5.1in}{\includegraphics[scale=0.1]{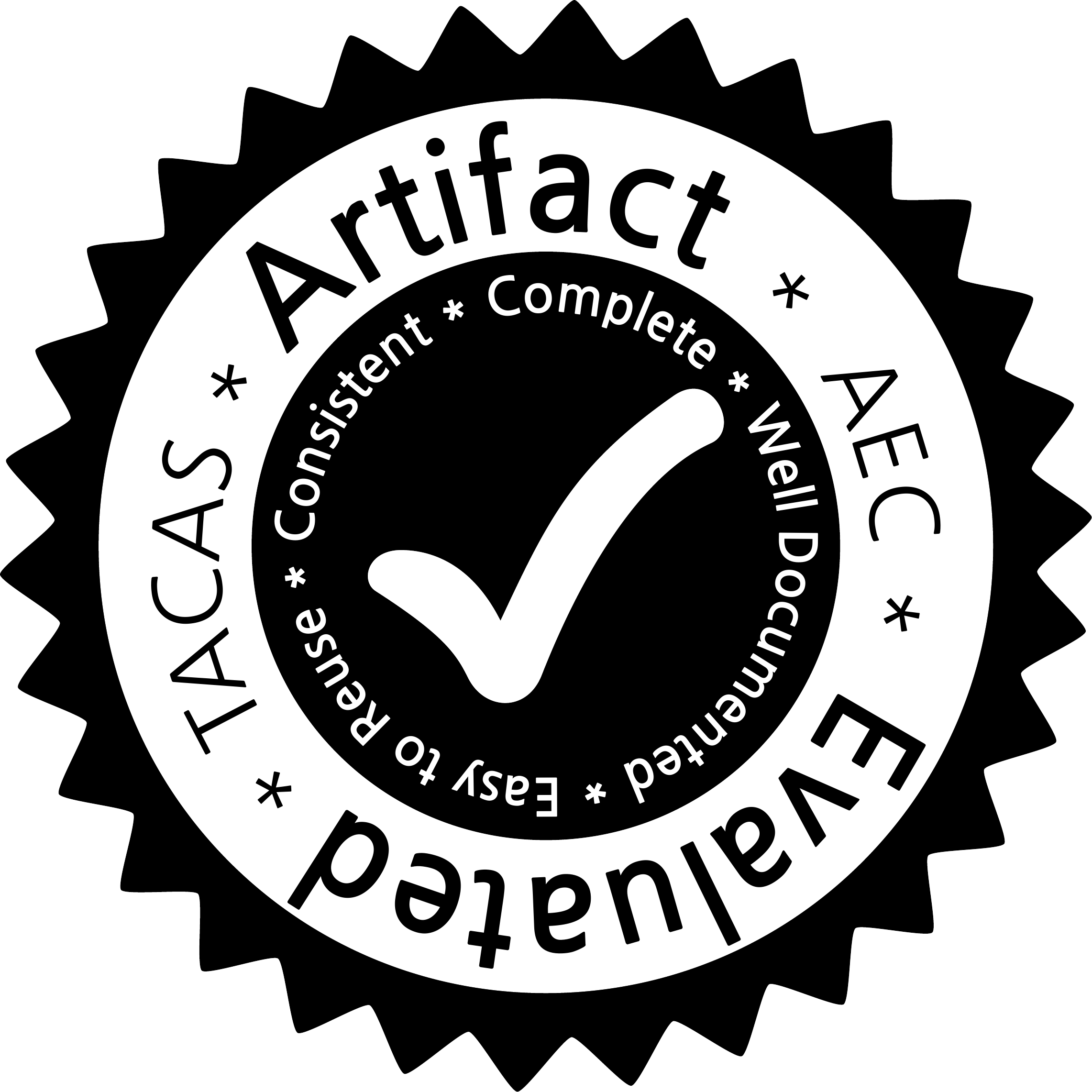}}}
\SetWatermarkAngle{0}

\title{\meta: Specification and Verification of High-Level Properties}
\author{
	Virgile~Robles \inst{1}\and
	Nikolai~Kosmatov\inst{1}\and
	Virgile~Prevosto\inst{1}\and
	Louis~Rilling\inst{2} \and
	Pascale~Le~Gall\inst{3}
}
\institute{Institut LIST, CEA, Université Paris-Saclay, Palaiseau, France\\
	\email{firstname.lastname@cea.fr}
	\and
	DGA, France, 
	\email{louis.rilling@irisa.fr}
	\and
	Laboratoire de Mathématiques et Informatique pour la Complexité et les Systèmes\\
	CentraleSupélec, Université Paris-Saclay, Gif-Sur-Yvette, France\\
	\email{pascale.legall@centralesupelec.fr}
}

\begin{document}
	\maketitle
\vspace{-5mm}	
	\begin{abstract}
          Modular deductive verification is a powerful technique
          capable to show that each function in a program satisfies
          its contract.  However, function contracts do not provide a
          global view of which high-level (e.g. security-related)
          properties of a whole software module are actually
          established, making it very difficult to assess them.
          To address this issue, this paper proposes a new
          specification mechanism, called meta-properties.  A
          meta-property can be seen as an enhanced global invariant
          specified for a set of functions, and capable to
          express predicates on values of variables, as well as memory
          related conditions (such as separation) and read or write
          access constraints.  We also propose an automatic
          transformation technique translating meta-properties into
          usual contracts and assertions, that can be proved by
          traditional deductive verification tools.  This technique
          has been implemented as a Frama-C plugin called MetAcsl and
          successfully applied to specify and prove safety- and
          security-related meta-properties in two illustrative case
          studies.

	\end{abstract}
\vspace{-5mm}	
	\setcounter{footnote}{0}
	\section{Introduction} \label{sec:intro}

Modular deductive verification is a well-known
technique\longv{~\cite{Hoare1969}} for formally proving that a program respects
some user-defined properties. It consists in providing for each function of the
program a \emph{contract}, which basically contains a \emph{precondition}
describing what the function expects from its callers, and a
\emph{postcondition} indicating what it guarantees when it successfully returns.
Logical formulas, known as \emph{verification conditions} or \emph{proof
obligations} (POs), can then be generated and given to automated theorem
provers. If all POs are validated, the body of the function fulfills its
contract.  Many deductive verification frameworks exist for various programming
and formal specification languages. We focus here on \framac~\cite{framac} and
its deductive verification plugin \Wp, which allows proving a C program correct
with respect to a formal specification expressed in ACSL~\cite{framac}.
	
However, encoding \emph{high-level} properties spanning across the entire
program in a set of Pre/Post-based contracts is not always immediate. In the
end, such high-level properties get split among many different clauses in
several contracts, without an explicit link between them.  Therefore, even if
each individual clause is formally proved, it might be very difficult for a
verification engineer, a code reviewer or a certification authority to convince
themselves that the provided contracts indeed ensure the expected high-level
properties.  Moreover, a software product frequently evolves  during
its lifetime, leading to numerous modifications in the code and
specifications.  Maintaining a high-level (e.g. security-related) property is
extremely complex without a suitable mechanism to formally specify and
automatically verify it after each update.

The purpose of the present work is to propose such a specification mechanism for
high-level properties, which we call \emph{meta-properties}, and to allow their
automatic verification on C code in  \framac thanks to a new plugin called
\meta.

\paragraph{Motivation}\label{sec:motivation} This work was motivated by several
previous projects.  During the verification of a hypervisor, we observed the
need for a mechanism of specification and automatic verification of high-level
properties, in particular, for global properties related to isolation and memory
separation.  Isolation properties are known as key properties in many
verification projects, in particular, for hypervisors and
micro-kernels~\cite{KleinTCS14}.

A similar need for specific high-level properties recently arose from a case
study on a confidentiality-oriented page management system submitted by an
industrial partner.  In this example, each page and each user (process) are
given a confidentiality level, and we wish to specify and verify that in
particular:
\begin{itemize}
	\item ($P_\mathrm{read}$) a user cannot read data from a page with a
		confidentiality level higher than its own;
	\item ($P_\mathrm{write}$) a user cannot write data to a page with a
		confidentialtiy level lower than its own.
\end{itemize}
This case study will be used as a running example in this paper.  As a second
case study (also verified, but not detailed in this paper), we consider a simple
smart house manager with several interesting properties such as: ``a door can
only be unlocked after a proper authentication or in case of alarm'' or
``whenever the alarm is ringing, all doors must be unlocked''.  Again, these
examples involve properties that are hard to express with function contracts
since they apply to the entire program rather than a specific
function.\footnote{ These examples are publicly available at
\url{https://huit.re/metacas}.}
\paragraph{Contributions}
The contributions of this paper\footnote{A longer version is available at
\url{https://arxiv.org/abs/1811.10509}} include:
	\begin{itemize}
		\item a new form of high-level properties, which we call
			\emph{meta-properties}, and an extension of the ACSL language able
			to express them (Sect.~\ref{sec:meta-spec}),
		\item a set of code transformations to translate meta-properties into 
			native ACSL annotations that can be proved via the usual
			methods (Sect.~\ref{sec:meta-verif}),
		\item a \framac plugin \meta able to parse C code
		    annotated with meta-properties and to perform 
		    the aforementioned code transformations (Sect.~\ref{sec:impl-results}),
		\item a case study: a confidentiality-oriented page system, where
			important security guarantees were expressed using meta-properties
			and automatically verified thanks to the code transformation with
			\meta (Sect.~\ref{sec:impl-results}).
	\end{itemize}

\section{Specification of Meta-properties}\label{sec:meta-spec}

	A meta-property is a property meant to express high-level requirements.
	As such, it is not attached to any particular function but instead to a set
	of functions. It is thus defined in the global scope and can only refer to
	global objects.
	
	To define a meta-property, the user must provide \textsf{(i)} the set of
	functions it will be applied to, \textsf{(ii)} a property (expressed in
	ACSL) and \textsf{(iii)} the \emph{context}, {\it i.e.} a characterization
	of the situations in which they want the property to hold in each of these
	functions (everywhere in the function, only at the beginning and the end,
	upon writing in a variable, etc.).  Furthermore, depending on the context,
	the property can refer to some special variables which we call
	\emph{meta-variables}.  Figure~\ref{fig:meta-examples} features a few
	examples of meta-properties further explained below.

	\begin{figure}[tb]
		\vspace{-20pt}
		\centering
		\begin{CACSL}[style=CACSLstyle]
struct Page { //Page handler structure
	char* data; //First address of the page
	enum allocation status; //ALLOCATED or FREE (ensured by $M_1$, lines 10-12)
	enum confidentiality level; /*Page level, CONFIDENTIAL or PUBLIC*/ }
enum confidentiality user_level; //Current user process level
struct Page metadata[PAGE_NB]; //All pages
struct Page* page_alloc(void); //Allocates a page
void page_read(struct Page*, char* buffer); //Reads a page
void page_encrypt(struct Page*); //Encrypts a page in place, makes it PUBLIC
/*@ meta $M_1$: \forall function f; \strong_invariant(f),
		\forall int page; 0 <= page < PAGE_NB ==>
		metadata[page].status == FREE || metadata[page].status == ALLOCATED;
	meta $M_2$: \forall function f; //Only page_encrypt can change levels of allocated pages
		! \subset(f, {page_encrypt}) ==> \writing(f),
		\forall int page; 0 <= page < PAGE_NB && metadata[page].status == ALLOCATED
		==> \separated(\written, &metadata[page].level);
	meta $M_3$: \forall function f; \reading(f), //Ensures $P_{\mathrm{read}}$
		\forall int page; 0 <= page < PAGE_NB && metadata[page].status == ALLOCATED
		&& user_level == PUBLIC && metadata[page].level == CONFIDENTIAL
		==> \separated(\read, metadata[page].data + (0 .. PAGE_LENGTH - 1));
*/ //Meta-property ensuring  $P_{\mathrm{write}}$ is defined similarly to $M_3$
		\end{CACSL}
		\vspace{-10pt}
		\caption{\protect\centering Partial meta-specification of a confidentiality case study}
		\label{fig:meta-examples}
	\end{figure}

	Let ${\mathcal{F}}$ denote the set of functions defined in the current program,
	and ${\mathcal P}$ the set of native ACSL properties.  Formally, we can define a
	meta-property as a triple $(c, F, P)$, where $c$ is a context (see
	Sect.~\ref{sec:context}), $F \subseteq {\mathcal{F}}$ and $P \in {\mathcal P}$.
	Intuitively, we can interpret this triple as ``$\forall f \in F$, $P$ holds for
	$f$ in the context $c$''.  For the meta-property to be well-formed, $P$ must be
	a property over a subset of $\mathcal G \cup \mathcal M(c)$, where $\mathcal G$
	is the set of variables available in the global scope of the program and
	$\mathcal M(c)$ is the set of meta-variables provided by the context $c$.

	\begin{sloppypar}
		The actual \meta syntax for defining a meta-property $(c, F, P)$ is
		\texttt{meta~[specification~of~F]~c,~P;} An example is given by property
		$M_1$ (cf. lines 10--12 in Figure~\ref{fig:meta-examples}), where $F =
		\mathcal{F}$, $c = \mathtt{strong\_invariant}$ and $P$ is the predicate
		stating that the status of any page should be either \verb|FREE| or
		\verb|ALLOCATED|.
	\end{sloppypar}

	\begin{subsection}{Target Functions and
		Quantification}\label{sec:quantif}

		Meta-properties are applied to a given \emph{target set} of functions
		$F$ defined as $F = F_+ \setminus F_-$ by providing explicit lists of
		considered and excluded functions $F_+, F_- \subseteq {\mathcal F}$.  If
		not provided, $F_+$ and $F_-$ are respectively equal to ${\mathcal F}$
		and $\emptyset$ by default, {\it i.e.} the meta-property should hold for
		all functions of the program. $F_-$ is useful when the user wants to
		target every function except a few, since they do not have to explicitly
		provide every resulting target function.

		\begin{sloppypar}
			The \meta syntax for the specification of $F$ uses the built-in ACSL
			construction \acsl{forall}, possibly followed by \acsl{subset} with
			or without logic negation  \lstinline|!| (to express $f\in F_+$ and
			$f\notin F_-$).  It can be observed in property $M_2$ (lines
			13--16),  where $F_+= \mathcal F$ and
			$F_-=\{\mathtt{page\_encrypt}\}$ excludes only one function.
		\end{sloppypar}
	\end{subsection}
	\begin{subsection}{Notion of Context}\label{sec:context}

		The \emph{context} $c$ of a meta-property defines the states in which
		property $P$ must hold, and may introduce \emph{meta-variables} that can
		be used in the definition of $P$.

		\paragraph{Beginning/Ending Context (Weak Invariant)} A \emph{weak
		invariant} indicates that  $P$ must hold at the beginning and at the end
		of each target function $f \in F$.

		\paragraph{Everywhere Context (Strong invariant)} A \emph{strong
		invariant} is similar to a weak invariant, except that it ensures that
		$P$ holds at \emph{every point}
		\footnote{
			More precisely, every \emph{sequence point} as defined by the C
			standard.
		}
		of each target function.  For example, property $M_1$ specifies that at
		every point of the program, the status of any page must be either
		\verb|FREE| or \verb|ALLOCATED|.

		\paragraph{Writing Context} This ensures that $P$ holds upon any
		modification of the memory (both stack and heap). It provides a
		meta-variable \acsl{written} that refers to the variable (and, more
		generally, the memory location) being written to.

		A simple usage of this context can be to forbid any direct modification
		of some global variable, as in property $M_2$.  This property states
		that for any function that is not \texttt{page\_encrypt}, the left-hand
		side of any assignment must be \emph{separated} from (that is, disjoint
		with) the global variable \texttt{metadata[page].level} for any
		\texttt{page} with the \verb|ALLOCATED| status. In other words, only the
		\verb|page_encrypt| function is allowed to modify the confidentiality
		level of an allocated page.

		An important benefit of this setting is a \emph{non-transitive
		restriction of modifications} that cannot be specified using the ACSL
		clause \texttt{assigns}, since the latter is transitive over function
		calls and necessarily permits to modify a variable when at least one
		callee has the right to modify it.  Here, since we only focus on
		\emph{direct} modifications, a call to \texttt{page_encrypt} (setting to
		public the level of the page it has encrypted) from another function
		does not violate  meta-property $M_2$.

		Furthermore, modification can be forbidden \emph{under some
		condition} (i.e. that the page is allocated), while
		\texttt{assigns} has no such mechanism readily available.
		
		\paragraph{Reading Context} Similar to the writing context, this ensures
		that the property holds whenever some memory location is read, and
		provides a  meta-variable \acsl{read} referring to the read location.
		It is used in property $M_3$ (lines 17--20), which expresses the
		guarantee $P_\mathrm{read}$ of the case study (see
		Sec.~\ref{sec:motivation}) by imposing a separation of a read location
		and the contents of allocated confidential pages when the user does not
		have sufficient access rights.  As another example, an isolation of a
		page can be specified as separation of all reads and writes from it.

		\smallskip

		These few simple contexts, combined with the native features of ACSL,
		turn out to be powerful enough to express quite interesting properties,
		including memory isolation and all properties used in our two motivating
		case studies.
	\end{subsection}

\section{Verification of Meta-properties}\label{sec:meta-verif}
	Figure~\ref{fig:meta-examples-impl} shows an (incorrect) toy implementation
	of two functions of Figure~\ref{fig:meta-examples} that we will use to
	illustrate the verification of meta-properties $M_1$--$M_3$.

	The key idea of the verification is the translation of meta-properties into
	native ACSL annotations, which are then verified using existing \framac
	analyzers. To that end, the property $P$ of a meta-property $(c, F, P)$ must
	be inserted as an assertion in relevant locations (as specified by context
	$c$) in each target function $f \in F$, and the meta-variables (if any) must
	be instantiated.
	
	\begin{wrapfigure}{r}{.5\textwidth}
		\vspace{-8mm}
		\centering
		\lstset{xleftmargin=1em,framexleftmargin=0.5em}
		\begin{CACSL}
			struct Page* page_alloc() {
				//try to find a free page
				struct Page* fp = find_free_page();
		    //if a free page is found, 
		    //allocate it with current user level
				if(fp != NULL) {
					fp->status = ALLOCATED;
					fp->level = user_level;
				} 
				return fp;
			}
			void page_read(struct Page* from, char* buffer) {
				for(unsigned i = 0 ; i < PAGE_LENGTH ; ++i)
					buffer[i] = from->data[i];
			}
		\end{CACSL}
		\vspace{-4mm}
		\caption{\protect\centering Incorrect code w.r.t. $M_2$ and
		$M_3$}
		\label{fig:meta-examples-impl}
		\vspace{-8mm}
	\end{wrapfigure}

	We define a specific translation for each context. For weak invariants, 
	property $P$ is simply added as both a precondition and a postcondition in the
	contract of $f$. This is also done for the strong invariant, for which 
	$P$ is additionally inserted after each instruction potentially modifying the
	values of the free variables in $P$\footnote{
		The AST is normalized so that every memory modification happens through
		an assignment. Then we conservatively determine if the object being
		assigned is one of the free variables of $P$: in presence of
		pointers, we assume the worst case.
	}
	For example, Figure~\ref{fig:meta-trans-m1} shows the translation of
	$M_1$ on \texttt{page\_alloc}. Our property (defined on lines 11--12 in
	Figure~\ref{fig:meta-examples}, denoted $P_{M_1}$ here) is inserted after
	the modification of a \verb|status| field (line 6) since the property
	involves these objects, but not after the modification of a \verb|level|
	field (line 8).

	For \emph{Writing} (resp. \emph{Reading}) contexts, $P$ is inserted before
	any instruction potentially making a write (resp. read) access to the
	memory, with the exception of function calls. In addition, each
	meta-variable is replaced by its actual value.  For example, in the
	translation of $M_2$ on \texttt{page\_alloc}
	(Figure~\ref{fig:meta-trans-m2}), the property is inserted before the two
	modifications of \verb|fp|, and \acsl{written} is replaced respectively by
	\verb|fp->status| and \verb|fp->level|. In this case $M_2$ does not hold.
	While its first instantiation (lines 4--6) is easily proved, it is not the
	case for the second one (lines 8--10). Indeed, there exists a \verb|page|
	(the one being modified) that has a status set to \lstinline|ALLOCATED|
	because of the previous instruction (line 7) and for which the
	\acsl{separated} clause is obviously false.  Hence, the assertion fails,
	meaning that the whole meta-property $M_2$ cannot be proved.  The fix
	consists in swapping lines 6 and 7 in Figure~\ref{fig:meta-examples-impl}.
	After that, all assertions generated from $M_2$ are proved.

	A similar transformation for $M_3$ on \verb|page_read| shows that the proof
	fails since the implementation allows an agent to read from any page without
	any check. Adding proper guards allows the meta-property to be proved.
	Conversely, if a meta-property is broken by an erroneous code update, a
	proof failure after \emph{automatically} re-running \meta helps to easily
	detect it.

	\longv{
		\smallskip
		This verification technique is related to~\cite{Pavlova2004} as we
		generate annotations whose verification implies the validity of a
		high-level property. However their specified properties are in the form
		of pre-/post-conditions on \emph{core} functions that are then
		propagated throughout the code, while we define properties that are not
		always pertaining to some core functions and use a simpler propagation
		method.
	}

	\begin{figure}[tb]
		\vspace{-6mm}
		\begin{subfigure}[c]{0.4\textwidth}
			\centering
			\begin{CACSL}
				/*@ requires P_M_1;
					ensures P_M_1; */
				struct Page* page_alloc() {
					struct Page* fp = find_free_page();
					if(fp != NULL) {
						fp->status = ALLOCATED;
						/*@ assert P_M_1;*/
						fp->level = user_level;
						//Line 8 cannot break P_M_1
					}
				}
			\end{CACSL}
			\vspace{-20pt}
			\caption{\centering Transformation for $M_1$}
			\label{fig:meta-trans-m1}
		\end{subfigure}
		\hspace{5mm}
		\begin{subfigure}[c]{0.6\textwidth}
			\centering
			\begin{CACSL}
				struct Page* page_alloc() {
					struct Page* fp = find_free_page();
					if(fp != NULL) {
						/*@ assert \forall int page; 0 <= page < PAGE_NB
							==> metadata[page].status == ALLOCATED
							==> \separated(fp->status, &metadata[page].level);*/
						fp->status = PAGE_ALLOCATED;
						/*@ assert \forall int page; 0 <= page < PAGE_NB
							==> metadata[page].status == ALLOCATED
							==> \separated(fp->level, &metadata[page].level);*/
						fp->level = user_level;
					}
				}
			\end{CACSL}
			\vspace{-30pt}
			\caption{\centering Transformation for $M_2$}
			\label{fig:meta-trans-m2}
		\end{subfigure}
		\caption{Examples of code transformations for functions of Figure~\ref{fig:meta-examples-impl}}
		\vspace{-1mm}
	\end{figure}

\vspace{-0.5mm}
\section{Results on Case Studies and Conclusion}\label{sec:impl-results}
\vspace{-0.5mm}

	\paragraph{Experiments} The support of meta-properties and the proposed
	methodology for their verification were fully implemented in OCaml as a
	\framac plugin  called \meta.  We realized a simple implementation of the
	two case studies mentioned in Sect.~\ref{sec:motivation}) and were able to
	fully specify and automatically verify all aforementioned properties (in
	particular $P_{\mathrm{read}}$ and $P_{\mathrm{write}}$) using \meta. The
	transformation step is performed in less than a second while the automatic
	proof takes generally less than a minute. 

	\paragraph{Conclusion} 
	We proposed a new specification mechanism for high-level properties in
	\framac, as well as an automatic transformation-based technique to verify
	these properties by a usual deductive verification approach.  \shortv{The
		main idea of this technique is similar to some previous efforts
	e.g.~\cite{Pavlova2004}}.  Meta-properties provide a useful extension to
	function contracts offering the possibility to express a variety of
	high-level safety- and security-related properties\shortv{.}\longv{,
	including e.g. memory isolation and confidentiality.}They also provide a
	verification engineer with an explicit global view of high-level properties
	being really proved, avoiding the risk to miss some part of an implicit
	property which is not formally linked to relevant parts of several function
	contracts, thus facilitating code review and certification. Another benefit
	of the new mechanism is the possibility to easily re-execute a proof after a
	code update\shortv{.}\longv{, diminishing the risk of making invalid an
	implicit high-level property spanned over several contracts.}Initial
	experiments confirm the interest of the proposed solution. 

	\paragraph{Future Work} 
	We plan to establish a formal soundness proof for our transformation
	technique, thereby allowing \meta to be reliably  used for critical code
	verification.  \longv{ Moreover, we wish to refine the transformation
	technique in order to allow the proof of the generated specification to
	scale better when the number of meta-properties or the size of the code
	increases.}Other future work directions include further experiments to
	evaluate the proposed approach on real-life software and for more complex
	properties.

	{\scriptsize \smallskip \noindent \textit{Acknowledgment}
		This work was partially supported by the project VESSEDIA, which has
		received funding from the EU Horizon 2020 research and innovation
		programme under grant agreement No 731453.  The work of the first author
		was partially funded by a Ph.D. grant of the French Ministry of Defense.
		Many thanks to the anonymous referees for their helpful comments.
	}

	\printbibliography{}
	\clearpage

\end{document}